\pdfoutput=1

\newif\ifdraft\draftfalse  
\newif\iffull\fullfalse   
\newif\ifbackref\backreffalse 

\makeatletter \@input{texdirectives.tex} \makeatother

\documentclass[sigplan,10pt]{acmart}

\setcopyright{none}             

\makeatletter
\fancypagestyle{firstpagestyle}{%
  \fancyhf{}%
  \renewcommand{\headrulewidth}{\z@}%
  \renewcommand{\footrulewidth}{\z@}%
    \fancyhead[L]{}%
    \fancyhead[R]{}%
}
\fancypagestyle{standardpagestyle}{%
  \fancyhf{}%
  \renewcommand{\headrulewidth}{\z@}%
  \renewcommand{\footrulewidth}{\z@}%
    \fancyhead[LE]{\@headfootfont\thepage}%
    \fancyhead[RO]{\@headfootfont\thepage}%
    \fancyhead[RE]{\@headfootfont\@shortauthors}%
    \fancyhead[LO]{\@headfootfont\shorttitle}%
    \fancyfoot[RO,LE]{}%
}
\def\@mkbibcitation{}
\makeatother

\renewcommand\footnotetextcopyrightpermission[1]{}

\usepackage[justification=centering]{caption}
\usepackage{afterpage}
\usepackage{color}
\usepackage{amssymb,amsmath,amsfonts}
\usepackage{extarrows}
\usepackage{version}
\usepackage{xspace}
\usepackage{graphicx}
\usepackage{listings}
\usepackage{wrapfig}
\usepackage{ifpdf}
\usepackage{semantic}
\usepackage{mathpartir} 
\usepackage{natbib}
\setcitestyle{numbers}
\usepackage{amsthm}
\usepackage{stmaryrd}
\usepackage{subfigure}
\usepackage{multirow}
\usepackage{proof}
\definecolor{darkblue}{rgb}{0.0,0.0,0.3}

\usepackage{color}
\usepackage{listings}
\usepackage[colorinlistoftodos]{todonotes}
\usepackage{tikz}
\usetikzlibrary{positioning,shadows,arrows,calc,backgrounds,fit,shapes,shapes.multipart,decorations.pathreplacing,shapes.misc,patterns}
\usepackage{tikzscale}
\usepackage{xspace}

\usepackage{hyperref}
\hypersetup{breaklinks=true}

\newcommand*{\EG}{e.g.,\xspace}
\newcommand*{\IE}{i.e.,\xspace}
\newcommand*{\ETAL}{et al.\xspace}

\begin{document}

\title{\huge Formally Secure Compilation of Unsafe Low-Level Components}

\subtitle{\vspace{-0.3em}(Extended Abstract)}

\author{
  Guglielmo Fachini\textsuperscript{1} \quad 
  C\u{a}t\u{a}lin Hri\c{t}cu\textsuperscript{1} \quad 
  Marco Stronati\textsuperscript{1} \quad 
  Ana Nora Evans\textsuperscript{1,2} \quad 
  Th\'eo Laurent\textsuperscript{1,3} \\[1ex] 
  Arthur Azevedo de Amorim\textsuperscript{4} \quad 
  Benjamin C. Pierce\textsuperscript{5} \quad 
  Andrew Tolmach\textsuperscript{6}\vspace{0.5em}} 
\affiliation{
  \textsuperscript{1}Inria Paris\quad
  \textsuperscript{2}University of Virginia\quad
  \textsuperscript{3}ENS Paris\quad
  \textsuperscript{4}Carnegie Mellon University\\
  \textsuperscript{5}University of Pennsylvania\quad
  \textsuperscript{6}Portland State University\vspace{0.5em}}

\makeatletter
\renewcommand{\@shortauthors}{Fachini~\ETAL}
\makeatother

\begin{abstract}
We propose a new formal criterion for secure compilation, providing strong
security guarantees for components written in unsafe, low-level languages
with C-style undefined behavior.  Our criterion goes beyond recent
proposals, which protect the trace properties of a single component against
an adversarial context, to model dynamic compromise in a system of mutually
distrustful components.  Each component is protected from all the others
until it receives an input that triggers an undefined behavior, causing it
to become compromised and attack the remaining uncompromised components.  To
illustrate this model, we demonstrate a secure compilation chain for an
unsafe language with buffers, procedures, and components, compiled to a
simple RISC abstract machine with built-in compartmentalization.  The
protection guarantees offered by this abstract machine can be achieved at
the machine-code level using either software fault isolation or tag-based
reference monitoring.  We are working on machine-checked proofs showing that
this compiler satisfies our secure compilation criterion.

\end{abstract}

\maketitle

\newcommand{\cmp}[1]{#1\hspace{-0.35em}\downarrow}

\newcommand{\citeFull}[2]{\iffull\cite{#1,#2}\else\cite{#1}}

\section*{Extended Abstract}
\label{sec:intro}

Computer systems are distressingly insecure.
Visiting a website, opening an email, or serving a client request often suffice
to open the door to control-hijacking attacks.
These devastating low-level attacks typically exploit memory-safety
vulnerabilities such as buffer overflows, use-after-frees, or double
frees, which are abundant in large software systems.

Various techniques have been proposed for guaranteeing memory
safety~\citeFull{NagarakatteMZ15, micropolicies2015}{
  NagarakatteZMZ09, NagarakatteZMZ10, DeviettiBMZ08, Nagarakatte2013,
  NagarakatteMZ14, interlocks_ahns2012, LowFat2013}, but
the challenges of
efficiency\iffull~\cite{NagarakatteZMZ09, NagarakatteZMZ10}\fi,
precision\iffull~\cite{m7}\fi,
scalability\iffull~\cite{ZitserLL04}\fi,
backwards compatibility\iffull~\cite{cheri_asplos2015}\fi, and
effective deployment\iffull~\cite{DeviettiBMZ08, Nagarakatte2013,
  NagarakatteMZ14, NagarakatteMZ15, interlocks_ahns2012,
  micropolicies2015, LowFat2013, pump_asplos2015}\fi{}
have hampered their widespread adoption.
Meanwhile, new mitigation techniques aim at dealing with the most
onerous consequences of memory unsafeness\iffull~\cite{AbadiBEL09}\fi.
In particular, {\em compartmentalization} offers a
strong, practical defense against low-level attacks
exploiting memory unsafeness~\cite{GudkaWACDLMNR15, cheri_oakland2015,
  wedge_nsdi2008}.
At least three compartmentalization technologies are
widely deployed: process-level privilege
separation~\cite{Kilpatrick03, GudkaWACDLMNR15, wedge_nsdi2008} (used,
\EG in OpenSSH~\cite{ProvosFH03} and for sandboxing plugins and tabs
in modern web browsers~\cite{ReisG09}), software fault
isolation~\cite{sfi_sosp1993} (provided, \EG by Google Native
Client~\cite{YeeSDCMOONF10}), and hardware enclaves (\EG Intel
SGX\iffull~\cite{sgx}\fi); many more are on the drawing
boards~\cite{micropolicies2015,
  cheri_oakland2015,PatrignaniDP16,cheri_asplos2015}.

Such low-level compartmentalization mechanisms are well suited for
building more secure compiler chains.
In particular, they can be exposed in unsafe low-level
languages like C and be targeted by their compiler chains to enable
efficiently breaking up large applications into mutually
distrustful components that run with minimal privileges and that can
interact only via well-defined interfaces.
Intuitively, protecting each component from all the others should have
strong security benefits: the compromise of some components
should not compromise the security of the whole application.

What, exactly, {\em are} the formal security guarantees one can obtain from
such secure compiler chains?
To answer this question, we start from {\em robust
  compilation}~\cite{GargHPSS17}, a recently proposed formal criterion
for secure compilation, which implies the preservation of all trace
properties even against adversarial contexts.
These traces are normally built over events such as inputs from and
outputs to the environment~\cite{Leroy09}.
We write $P \Downarrow t$ to mean that the complete program $P$ can
produce trace $t$ with respect to some operational semantics.
Armed with this, robust compilation is formally stated as:
\[
\forall P~C_T~t.~
  C_T[\cmp{P}] \Downarrow t \Rightarrow  \exists C_S.~ C_S[P] \Downarrow t
\]
For any partial source program $P$ and any (adversarial) target
context $C_T$ where $C_T$ linked with the compiled variant of $P$
can produce a (bad) trace $t$ in the target language
(written $C_T[\cmp{P}] \Downarrow t$), we can construct a(n adversarial)
source-level context $C_S$ that can produce trace $t$ in the source
language when linked with $P$ (\IE $ C_S[P] \Downarrow t$).
Intuitively, any attack trace $t$ that context $C_T$ can mount against
$\cmp{P}$ can already be mounted against $P$ by some source language
context $C_S$.
Conversely, any trace property that holds of
$P$ when linked with any arbitrary source context will still hold for $\cmp{P}$
when linked with an arbitrary target context.

In this work, we propose a new formal criterion for secure compilation
that extends robust compilation to protecting mutually distrustful
components against each other in an unsafe low-level language with
C-style undefined behavior.
The characterization of robust compilation above does not directly apply
in this setting, since it assumes the source language is safe and $P$
cannot have undefined behavior.
The natural way to adapt robust compilation to an unsafe source
language is the following:
\[
\forall P~C_T~t.~
  C_T[\cmp{P}] \Downarrow t \Rightarrow  \exists C_S~t'.~ C_S[P] \Downarrow t'
  \wedge t' \preccurlyeq_P t
\]
Instead of requiring that $C_S[P]$ perform the entire trace $t$,
we also allow it to produce a finite prefix $t'$ that ends with an
undefined behavior in $P$ (which we write as $t' \preccurlyeq_P t$).
Intuitively, since we want to reason only in terms of safe source
contexts we do not allow $C_S$ to exhibit undefined behaviors.
However, even a safe context can sometimes trigger an undefined
behavior in the protected program $P$, in which case there is no way
to keep protecting $P$ going forward.
However, $P$ is fully protected until it receives an
input that causes undefined behavior.
This is a good step towards a model of {\em dynamic compromise}.

We show that this can be extended to support
{\em mutual distrustful components}.
We start by taking both partial programs and contexts to be sets of
components and plugging a program in a context to be linking.
We compile sets of components by separately compiling each component.
We start with all components being uncompromised and incrementally
replace any component that exhibits undefined behavior in the source
with an arbitrary safe component that will now attack the
remaining uncompromised components.
Formally, this is captured by the following property:
\[
\begin{array}{l}
\cmp{\{C_1,...,C_n\}} ~\Downarrow~ t \Rightarrow \exists A_{i_1},...,A_{i_m}.\\
\quad (1)~ (\{C_1,...,C_n\} \backslash \{C_{i_1},...,C_{i_m}\} \cup \{A_{i_1},...,A_{i_m}\}) ~\Downarrow~ t\\
\quad (2)~\forall j \in 1 ... m.~ \exists t' \prec_{C_{i_j}} t.~\\
\quad\qquad  \cmp{(\{C_1,...,C_n\} \backslash \{C_{i_1},...,C_{i_j{-}1}\} \cup \{A_{i_1},...,A_{i_j{-}1}\})} ~\Downarrow t'
\end{array}
\]
What this is saying is that each low-level trace $t$ of a compiled set of
components $\cmp{\{C_1,...,C_n\}}$ can be reproduced in the source
language after replacing the compromised components $\{C_{i_1},...,C_{i_m}\}$
with source components $\{A_{i_1},...,A_{i_m}\}$ (part $1$).
Moreover (part $2$), $C_{i_1},...,C_{i_m}$ precisely
characterizes a compromise sequence, in which each component $C_{i_j}$
is taken over by the already compromised components at that point in
time $\{A_{i_1},...,A_{i_j{-}1}\}$.
Above we write $t' \prec_{C_{i_j}} t$ for expressing that
trace $t'$ is a prefix of $t$ ending with an undefined behavior
in $C_{i_j}$.

This new definition allows us to play an iterative game in which each
component is protected until it receives an input that triggers an
undefined behavior, causing it to become compromised and to attack the
remaining uncompromised components.
This is the first security definition in this space to support both
dynamic compromise and mutual distrust, whose interaction is subtle
and has eluded previous attempts at characterizing the security
guarantees of compartmentalizing compilation as extensions of fully
abstract compilation~\cite{JuglaretHAEP16}.
We further show that this new security definition can be obtained by
repeatedly applying an instance of the simpler robust compilation
definition above that is phrased only in terms of a program and a
context.

We illustrate this new security definition by applying
it to a new secure compilation chain for an unsafe language
with buffers, procedures, components, and a CompCert-inspired memory
model~\cite{LeroyB08}.
We compile this language to a simple RISC abstract machine with
built-in compartmentalization and are working on constructing
machine-checked Coq proofs that this compiler satisfies our secure
compilation definition.
In terms of proof effort, we reduce (a safety-variant of) our
property to providing a back-translation of individual finite trace
prefixes to safe high-level contexts together with three standard
simulation proofs, while previous proofs in this space were much more
complex and non-standard~\cite{JuglaretHAEP16}.
This gives us hope that our security definition and proof technique
can be scaled to something as large as a
secure variant of CompCert.

Finally, the protection of the compartmentalized abstract
machine can be achieved at the lowest machine-code level using either
software-fault isolation~\cite{sfi_sosp1993} or tag-based reference
monitoring~\cite{micropolicies2015}.
We designed and are building two such back ends for our compiler,
using property-based testing to validate that the two are functional
and secure.

\bibliographystyle{plainurl}
\bibliography{mp,safe}

\begin{thebibliography}{\tiny10}

\bibitem{micropolicies2015}
A.~{Azevedo de Amorim}, M.~D\'en\`es, N.~Giannarakis, C.~Hri\c{t}cu, B.~C.
  Pierce, A.~{Spector-Zabusky}, and A.~Tolmach.
\newblock
  \href{http://prosecco.gforge.inria.fr/personal/hritcu/publications/micro-policies.pdf}{Micro-policies:
  Formally verified, tag-based security monitors}.
\newblock \iffull{In {\em 36th IEEE Symposium on Security and Privacy (Oakland
  S\&P)}}\else{{\em Oakland S\&P}}\fi{}. 2015.

\bibitem{wedge_nsdi2008}
A.~Bittau, P.~Marchenko, M.~Handley, and B.~Karp.
\newblock
  \href{http://www.usenix.org/legacy/events/nsdi08/tech/full_papers/bittau/bittau.pdf}{Wedge:
  Splitting applications into reduced-privilege compartments}.
\newblock \iffull{In {\em USENIX Symposium on Networked Systems Design and
  Implementation}}\else{{\em USENIX NSDI}}\fi{}, 2008.

\bibitem{cheri_asplos2015}
D.~Chisnall, C.~Rothwell, R.~N.~M. Watson, J.~Woodruff, M.~Vadera, S.~W. Moore,
  M.~Roe, B.~Davis, and P.~G. Neumann.
\newblock
  \href{https://www.cl.cam.ac.uk/~dc552/papers/asplos15-memory-safe-c.pdf}{Beyond
  the {PDP-11}: Architectural support for a memory-safe {C} abstract machine}.
\newblock \iffull{In {\em Proceedings of the Twentieth International Conference
  on Architectural Support for Programming Languages and Operating
  Systems}}\else{{\em ASPLOS}}\fi{}. 2015.

\bibitem{GargHPSS17}
D.~Garg, C.~Hri\c{t}cu, M.~Patrignani, M.~Stronati, and D.~Swasey.
\newblock \href{https://arxiv.org/abs/1710.07309}{Robust hyperproperty
  preservation for secure compilation (extended abstract)}.
\newblock arXiv:1710.07309, 2017.

\bibitem{GudkaWACDLMNR15}
K.~Gudka, R.~N.~M. Watson, J.~Anderson, D.~Chisnall, B.~Davis, B.~Laurie,
  I.~Marinos, P.~G. Neumann, and A.~Richardson.
\newblock \href{https://www.cl.cam.ac.uk/~kg365/pubs/2015ccs-soaap.pdf}{Clean
  application compartmentalization with {SOAAP}}.
\newblock \iffull{In {\em 22nd {ACM} {SIGSAC} Conference on Computer and
  Communications Security}}\else{{\em CCS}}\fi{}. 2015.

\bibitem{JuglaretHAEP16}
Y.~Juglaret, C.~Hritcu, A.~{Azevedo de Amorim}, B.~Eng, and B.~C. Pierce.
\newblock \href{https://doi.org/10.1109/CSF.2016.11}{Beyond good and evil:
  Formalizing the security guarantees of compartmentalizing compilation}.
\newblock \iffull{In {\em {IEEE} 29th Computer Security Foundations Symposium,
  {CSF} 2016, Lisbon, Portugal, June 27 - July 1, 2016}}\else{{\em CSF}}\fi{},
  2016.

\bibitem{Kilpatrick03}
D.~Kilpatrick.
\newblock
  \href{http://www.usenix.org/events/usenix03/tech/freenix03/kilpatrick.html}{Privman:
  {A} library for partitioning applications}.
\newblock \iffull{In {\em Proceedings of the {FREENIX} Track: 2003 {USENIX}
  Annual Technical Conference, June 9-14, 2003, San Antonio, Texas,
  {USA}}}\else{{\em USENIX FREENIX}}\fi{}. 2003.

\bibitem{Leroy09}
X.~Leroy.
\newblock \href{http://dx.doi.org/10.1007/s10817-009-9155-4}{A formally
  verified compiler back-end}.
\newblock \iffull{{\em Journal of Automated Reasoning}}\else{{\em JAR}}\fi{},
  43(4):363--446, 2009.

\bibitem{LeroyB08}
X.~Leroy and S.~Blazy.
\newblock
  \href{http://pauillac.inria.fr/~xleroy/publi/memory-model-journal.pdf}{Formal
  verification of a {C}-like memory model and its uses for verifying program
  transformations}.
\newblock \iffull{{\em Journal of Automated Reasoning}}\else{{\em JAR}}\fi{},
  41(1):1--31, 2008.

\bibitem{NagarakatteMZ15}
S.~Nagarakatte, M.~M.~K. Martin, and S.~Zdancewic.
\newblock \href{http://drops.dagstuhl.de/opus/volltexte/2015/5026/}{Everything
  you want to know about pointer-based checking}.
\newblock \iffull{In {\em 1st Summit on Advances in Programming
  Languages}}\else{{\em SNAPL}}\fi{}. 2015.

\bibitem{PatrignaniDP16}
M.~Patrignani, D.~Devriese, and F.~Piessens.
\newblock \href{http://arxiv.org/abs/1604.05044}{On modular and fully-abstract
  compilation}.
\newblock \iffull{In {\em 29th IEEE Computer Security Foundations
  Symposium}}\else{{\em CSF}}\fi{}, 2016.

\bibitem{ProvosFH03}
N.~Provos, M.~Friedl, and P.~Honeyman.
\newblock
  \href{https://www.usenix.org/conference/12th-usenix-security-symposium/preventing-privilege-escalation}{Preventing
  privilege escalation}.
\newblock In {\em 12th {USENIX} Security Symposium}. 2003.

\bibitem{ReisG09}
C.~Reis and S.~D. Gribble.
\newblock
  \href{https://homes.cs.washington.edu/~gribble/papers/eurosys-2009.pdf}{Isolating
  web programs in modern browser architectures}.
\newblock \iffull{In {\em EuroSys Conference}}\else{{\em EuroSys}}\fi{}. 2009.

\bibitem{sfi_sosp1993}
R.~Wahbe, S.~Lucco, T.~E. Anderson, and S.~L. Graham.
\newblock
  \href{http://www.eecs.harvard.edu/~greg/cs255sp2004/wahbe93efficient.pdf}{Efficient
  software-based fault isolation}.
\newblock \iffull{In {\em Proceedings of the Symposium on Operating Systems
  Principles}}\else{{\em SOSP}}\fi{}, 1993.

\bibitem{cheri_oakland2015}
R.~N.~M. Watson, J.~Woodruff, P.~G. Neumann, S.~W. Moore, J.~Anderson,
  D.~Chisnall, N.~H. Dave, B.~Davis, K.~Gudka, B.~Laurie, S.~J. Murdoch,
  R.~Norton, M.~Roe, S.~Son, and M.~Vadera.
\newblock \href{http://dx.doi.org/10.1109/SP.2015.9}{{CHERI:} {A} hybrid
  capability-system architecture for scalable software compartmentalization}.
\newblock \iffull{In {\em 2015 {IEEE} Symposium on Security and Privacy, {SP}
  2015, San Jose, CA, USA, May 17-21, 2015}}\else{{\em IEEE S\&P}}\fi{}, 2015.

\bibitem{YeeSDCMOONF10}
B.~Yee, D.~Sehr, G.~Dardyk, J.~B. Chen, R.~Muth, T.~Ormandy, S.~Okasaka,
  N.~Narula, and N.~Fullagar.
\newblock \href{http://research.google.com/pubs/archive/34913.pdf}{{Native
  Client}: a sandbox for portable, untrusted x86 native code}.
\newblock \iffull{{\em Communications of the ACM}}\else{{\em CACM}}\fi{},
  53(1):91--99, 2010.

\end{thebibliography}

\end{document}